\documentstyle[amsmath,a4,12pt,epsfig]{article}
\def\slash{\!\!\!\!/}
\def\slash{\!\!\!/}

\begin{document}     
\title{{\bf A heavy quark effective field lagrangian keeping particle
and antiparticle mixed sectors}
\vspace*{0.2in}
\thanks{Research partially supported by CICYT under contract AEN-96/1718}}
\author{F. Bert\'o, J.L. Domenech and M.A. Sanchis-Lozano 
\thanks{Corresponding author; E-mail: mas@evalo1.ific.uv.es} \\
\\
\it Departamento de F\'{\i}sica Te\'orica and IFIC\\
\it Centro Mixto Universitat de Val\`encia-CSIC\\
\it 46100 Burjassot, Valencia, Spain}
\maketitle 
\abstract{We derive a tree-level heavy quark effective Lagrangian 
keeping particle-antiparticle mixed sectors allowing for
heavy quark-antiquark pair annihilation and
creation. However, when removing the unwanted degrees of freedom
from the effective Lagrangian one has to be careful in using the classical 
equations of motion obeyed by the effective fields in order to get a 
convergent expansion on the reciprocal of the heavy quark mass. 
Then the application of the effective theory to such hard processes
should  be sensible for special kinematic regimes as for example 
heavy quark pair production near threshold.}
\vspace{-15.cm}
\begin{flushright}
  FTUV/98-80 (revised)\\
  IFIC/98-81 (revised)
\end{flushright} 
\vspace{15.cm}
\begin{small}
\end{small}
\newpage
\section{Introduction}
\vspace{0.1in}
The present decade has witnessed the rapid development of an effective theory 
for the strong interaction (HQET), successfully applied to 
the phenomenology of hadrons containing a heavy quark 
\cite{grins,neub,mannel}.  Although without providing by itself 
$\lq\lq$solutions" to the non-perturbative
dynamics of QCD, HQET shows the existence of the celebrated spin-flavor
symmetry in the infinite quark mass limit \cite{isgur}, allowing a systematic
incorporation of finite mass symmetry breaking effects, obtaining
relations between physical observables and simplifying some
calculations.
\par 
However, some confusion regarding HQET as a quantum
field theory may appear in reading some classical papers on 
this topic \cite{neub} and its sequels \cite{neub2,pich}. In particular the
interpretation of the heavy-quark effective fields in terms of
creation/annihilation operators may be somewhat
misleading in those references and has to be clarified. A first attempt was 
done by one of us (M.A.S.L.) in \cite{mas97} where fields were Fourier
expanded as plane-wave components, thereby deriving the
free-particle Feynman propagators for the effective static theory. 
\par

In this paper we keep both heavy quark and heavy antiquark coupled sectors
in the HQET Lagrangian. To our knowledge, only the paper by
Wu \cite{wu} in the literature actually deals with both heavy
quark and antiquark fields altogether in the effective Lagrangian 
\footnote{Although we do not understand completely his final
expression (14) in \cite{wu}}. In the present work, we pursue this line
of investigation further, looking for more symmetric
expressions and a more transparent physical
\vspace{0.1in} interpretation. \par

\section{A complete tree-level HQET Lagrangian} 
\vspace{0.1in}
Below we firstly introduce the notation and some
definitions for the heavy-quark effective fields. Although well-known and
widely used in the literature, some remarks will be in order for
our later development. In subsection 2.2 we actually get started by
expressing the Lagrangian in terms of the effective fields 
keeping all non-null terms, leading in principle 
to the possibility of annihilation or creation of heavy quark-antiquark
 pairs.
\par
The heavy-quark effective theory is applicable for
almost on-shell heavy quarks. Introducing the 
$\lq\lq$residual" momentum $k$ for a heavy quark of momentum $p$ as
\footnote{Note that in a plane wave Fourier expansion of fields
$k$ should be properly named Fourier residual momentum, whose 
spatial components range 
from $-{\infty}$ to $+{\infty}$ before imposing any cut on the Fourier
expansion of fields. Once constructed the effective theory only 
the low-energy modes remain, i.e. $\vec{k}$ components of order of
${\Lambda}_{QCD}$ 
\cite{mas97,capitani}.}
\[ p\ =\ mv\ +\ k \]
the nearly on-shellness condition is written as
\begin{equation}
v{\cdot}k\ {\simeq}\ -\frac{k^2}{2m}
\end{equation}
but not necessarily $v{\cdot}k=0$ (only so in the infinite quark mass
limit).
Note also that in a reference frame moving with velocity $v$, the 
above condition is equivalent to
\begin{equation}
k^0\ =\ \sqrt{m^2+\vec{k}^2}\ - m\ \ {\simeq}\ \ \frac{\vec{k}^2}{2m}
\end{equation}
that is, $k^0$ can be identified with the heavy-quark kinetic energy (and its
non-relativistic limit).
\par 
The underlying idea when introducing the residual momentum is 
that once removed the large mechanical momentum associated to the 
heavy-quark mass, only the low-energy modes remain in the effective theory.
This is the standard way to handle heavy quarks in singly heavy
hadrons according to HQET. However, one 
may conjecture about the possibility of performing such an
energy-momentum shift by introducing a center-of-mass residual
momentum for processes involving creation or annihilation
of heavy quark-antiquark pairs at tree-level. In this paper we shall 
study the latter processes in the light of an effective QCD Lagrangian
instead of directly writing currents as bilinears 
mixing both particle and antiparticle sectors as in
\vspace{0.1in} \cite{schuler}.

\subsection{Effective fields}
\vspace{0.1in}
Following the standard reference \cite{neub}, we shall
introduce the effective fields for a heavy quark bound inside a
hadron moving with (four-)velocity $v$, as  
\begin{equation}
h_v^{(+)}(x)\ =\ e^{imv{\cdot}x}\ \frac{1+v{\slash}}{2}\ Q^{(+)}(x)
\end{equation}
\begin{equation}
H_v^{(+)}(x)\ =\ e^{imv{\cdot}x}\ \frac{1-v{\slash}}{2}\ Q^{(+)}(x)
\end{equation} 
where $Q^{(+)}(x)$ stands for the (positive energy) fermionic field describing 
the heavy quark in the full theory (see footnote $\#4$); $h_v^{(+)}(x)$
and $H_v^{(+)}(x)$ represent the $\lq\lq$large'' and $\lq\lq$small''
 components of a classical spinor field respectively.\par
Similarly for a heavy antiquark,
\begin{equation}
h_v^{(-)}(x)\ =\ e^{-imv{\cdot}x}\ \frac{1-v{\slash}}{2}\ Q^{(-)}(x)
\end{equation}
\begin{equation}
H_v^{(-)}(x)\ =\ e^{-imv{\cdot}x}\ \frac{1+v{\slash}}{2}\ Q^{(-)}(x)
\end{equation}
where $Q^{(-)}(x)$ stands for the (negative energy) anti-fermionic 
field in the full theory. 
\par
Therefore orthogonality implies (see appendix)
\[
\overline{h}_v^{(-)}h_v^{(+)}\ =\ \overline{h}_v^{(-)}H_v^{(-)}\ =\ 
\overline{H}_v^{(-)}h_v^{(-)}\ =\
\overline{H}_v^{(-)}H_v^{(+)}\ =\ 0
\]
but
\[
\overline{h}_v^{(-)}H_v^{(+)}\ {\neq}\ 0\ \ ;\ 
\overline{h}_v^{(-)}{\gamma}^{\mu}\ h_v^{(+)}\ {\neq}\ 0 
\]
and so on.\par

It can be sometimes read in
the literature \cite{neub,pich} that $h_v^{(+)}$ annihilates a heavy quark 
whereas allegedly $H_v^{(+)}$ creates a heavy antiquark, both with velocity 
$v$. Although $H_v^{(+)}$ indeed takes into account the antiparticle
features of a relativistic fermion, its Fourier decomposition \cite{mas97}
reads

\begin{equation} 
H_v^{(+)}(x)\ =\ \frac{1-v{\slash}}{2}\ \int\
\frac{d^3\vec{k}}{J}\ \sum_{r}\ b_r(\vec{k})\ 
u_r(\vec{k})\ e^{-ik{\cdot}x} \newline 
\end{equation} 
where the integral runs over small $\vec{k}$ components; 
$J$ is a kinematic factor depending on the
normalization chosen for the states \cite{mas97,korner2} and $b_r$ stands for
an annihilation operator of a heavy quark with spin denoted by $r$.
Let us stress that $H_v^{(+)}$ contains only annihilation
operators like \vspace{0.2in} $h_v^{(+)}$. 
\subsection{Derivation of the effective Lagrangian}
\vspace{0.1in}
In contrast to conventional low-energy effective theories in which
heavy fields are completely integrated out, in HQET 
the degrees of freedom associated with heavy quarks must not
be entirely removed, as one wants to describe real hadrons
containing them. Only the $\lq\lq$small" components of fields have to be 
removed, for example employing to this end
the classical equations of motion, according to the procedure used in 
\cite{neub}. 
\par
The main difference of this paper w.r.t. other standard works is 
that we are concerned with the existence of terms in the
effective Lagrangian mixing large components of the 
heavy quark and heavy antiquark fields, 
i.e. $h_v^{({\pm})}{\Gamma}h_v^{({\mp})}$, 
where ${\Gamma}$ stands for a combination of Dirac gamma matrices
and covariant derivatives.
\par

\def\slash{\!\!\!\!/}

The tree-level QCD Lagrangian is our point of departure: 
\footnote{Throughout this paper we keep upper arrows explicitly
indicating the action of derivatives} 
\begin{equation}
{\cal L}\ =\ \overline{Q}\ (i\overrightarrow{D{\slash}}-m)\ Q
\end{equation}
where
\begin{equation}
Q\ =
\ Q^{(+)}\ +\ Q^{(-)}\ =\ e^{-imv{\cdot}x}\biggl[\ h_v^{(+)}+H_v^{(+)}\ 
\biggr]\ +\ 
e^{imv{\cdot}x}\ \biggl[\ h_v^{(-)}+H_v^{(-)}\ \biggr]
\end{equation}
and $D$ standing for the covariant derivative
\begin{equation}
\overrightarrow{D}^{\mu}\ =\ \overrightarrow{\partial}^{\mu}\ -\ igT_aA_a^{\mu}
\end{equation}
with $T_a$ the generators of $SU(3)_c$. Substituting (9) in (8)
one easily arrives at
\begin{equation}
{\cal L}\ =\ {\cal L}^{(++)}\ +\ {\cal L}^{(--)}\ +\ {\cal L}^{(-+)}\ +\ 
{\cal L}^{(+-)}
\end{equation}
where we have explicitly splitted the Lagrangian into four different
pieces corresponding to the particle-particle, antiparticle-antiparticle
and both particle-antiparticle sectors. The former one has the form   
\begin{equation}
{\cal L}^{(++)}\ =\ \overline{h}_v^{(+)}(iv{\cdot}\overrightarrow{D})h_v^{(+)}-
\overline{H}_v^{(+)}(iv{\cdot}\overrightarrow{D}+2m)H_v^{(+)}+
\overline{h}_v^{(+)}i\overrightarrow{D{\slash}}_{\bot}H_v^{(+)}+
\overline{H}_v^{(+)}i\overrightarrow{D{\slash}}_{\bot}h_v^{(+)}
\vspace{0.1in}
\end{equation}
corresponding to the usual effective Lagrangian still containing
both $h_v^{(+)}$ and $H_v^{(+)}$ $\lq\lq$large" and $\lq\lq$small"
fields respectively. We employ the common notation where
perpendicular indices are implied according to
\[ D^{\mu}_{\bot}\ =\ D_{\alpha}\ (g^{\mu\alpha}-v^{\mu}v^{\alpha}) \]
\par
Regarding the antiquark sector of the theory
\begin{equation}
{\cal L}^{(--)}\ =\ -\overline{h}_v^{(-)}(iv{\cdot}\overrightarrow{D})
h_v^{(-)}+
\overline{H}_v^{(-)}(iv{\cdot}\overrightarrow{D}-2m)H_v^{(-)}+
\overline{h}_v^{(-)}i\overrightarrow{D{\slash}}_{\bot}H_v^{(-)}+
\overline{H}_v^{(-)}i\overrightarrow{D{\slash}}_{\bot}h_v^{(-)}
\end{equation}
\par
The latter expressions (12) and (13) considered as
quantum field Lagrangians, do not afford tree-level
heavy quark-antiquark pair
creation or annihilation processes stemming from the terms
mixing ${h}_v^{(\pm)}$ and ${H}_v^{(\pm)}$ fields since they contain
either annihilation and creation operators of heavy quarks
or annihilation and creation operators of heavy antiquarks
separately.\par
Nevertheless there are two extra pieces in the Lagrangian (11):
\begin{equation}
{\cal L}^{(-+)}=e^{-i2mv{\cdot}x}\ 
[\overline{H}_v^{(-)}(iv{\cdot}\overrightarrow{D})h_v^{(+)}
-\overline{h}_v^{(-)}(iv{\cdot}\overrightarrow{D}+2m)H_v^{(+)}+
\overline{h}_v^{(-)}i\overrightarrow{D{\slash}}_{\bot}h_v^{(+)}+
\overline{H}_v^{(-)}i\overrightarrow{D{\slash}}_{\bot}H_v^{(+)}]  
\vspace{0.1in}
\end{equation}
and
\begin{equation}
{\cal L}^{(+-)}=e^{i2mv{\cdot}x}\ 
[-\overline{H}_v^{(+)}(iv{\cdot}\overrightarrow{D})h_v^{(-)}
+\overline{h}_v^{(+)}(iv{\cdot}\overrightarrow{D}-2m)H_v^{(-)}+
\overline{h}_v^{(+)}i\overrightarrow{D{\slash}}_{\bot}h_v^{(-)}+
\overline{H}_v^{(+)}i\overrightarrow{D{\slash}}_{\bot}H_v^{(-)}] 
\vspace{0.1in}
\end{equation}
where use was made of the orthogonality of the $h_v^{(\pm)}$ and 
$H_v^{(\pm)}$ fields.
As could be expected, there are indeed pieces mixing both 
quark and antiquark fields
leading to the possibility of annihilation/creation processes 
deriving directly from the HQET Lagrangian. After all the Lagrangian (11) is
still equivalent to full (tree-level) QCD. 
\par

The Lagrangian (11) as developed in Eqs. (12-15) has too many (fermionic)
degrees of freedom if $h_v^{({\pm})}$ and  $H_v^{({\pm})}$ are
considered as independent fields. Thereby Neubert \cite{neub}
uses the first of the classical equations of motion, namely
\begin{eqnarray}
& & i\overrightarrow{D{\slash}}_{\bot}h_v^{({\pm})}\ -\ 
({\pm}iv{\cdot}\overrightarrow{D}+2m)\ H_v^{({\pm})}\ 
=\ 0 \\
& & {\pm}iv{\cdot}\overrightarrow{D}\ h_v^{({\pm})}\ +\ 
i\overrightarrow{D}{\slash}_{\bot}H_v^{({\pm})}\ =\ 0
\end{eqnarray}
to eliminate the unwanted degrees of freedom associated to
the small components $H_v^{({\pm})}$. Let us emphasize that
the above expressions consist of
two coupled equations for the particle sector (linking
$h_v^{(+)}$ and  $H_v^{(+)}$) and two coupled
equations for the antiparticle sector (linking
$h_v^{(-)}$ and  $H_v^{(-)}$). As expected if both
equations (16-17) were simultaneously used all Lagrangians
(12-15) would identically vanish.\par
\par
From Eq. (16) we may write
\begin{eqnarray}
H_v^{(+)} & = & (iv{\cdot}\overrightarrow{D}+2m-i\epsilon)^{-1}
i\overrightarrow{D{\slash}}_{\bot}h_v^{(+)} \\
H_v^{(-)} & = & (-iv{\cdot}\overrightarrow{D}+2m-i\epsilon)^{-1}
i\overrightarrow{D{\slash}}_{\bot}h_v^{(-)}
\end{eqnarray}
and for the conjugate fields
\begin{eqnarray}
\overline{H}_v^{(+)} & = & 
\overline{h}_v^{(+)}i\overleftarrow{D}{\slash}_{\bot}
(iv{\cdot}\overleftarrow{D}-2m+i\epsilon)^{-1} \\
\overline{H}_v^{(-)} & = & 
\overline{h}_v^{(-)}i\overleftarrow{D}{\slash}_{\bot}
(-iv{\cdot}\overleftarrow{D}-2m+i\epsilon)^{-1}
\end{eqnarray}
where

\[ \overleftarrow{D}^{\mu}\ =\ \overleftarrow{\partial}^{\mu}\ +\ igT_aA_a^{\mu} \]

\par
Substituting Eqs. (18-21) into (14-17), we get the following
non-local Lagrangians associated with the different sectors of the theory
\begin{eqnarray}
{\cal L}^{(++)} & = & \ 
\overline{h}_v^{(+)}(iv{\cdot}\overrightarrow{D})h_v^{(+)}\ +\ 
\overline{h}_v^{(+)}i\overrightarrow{D{\slash}}_{\bot}
(iv{\cdot}\overrightarrow{D}+2m)^{-1}
i\overrightarrow{D{\slash}}_{\bot}h_v^{(+)} \\
& & \nonumber \\
{\cal L}^{(--)} & = & \ 
\overline{h}_v^{(-)}(-iv{\cdot}\overrightarrow{D})h_v^{(-)}-\ 
\overline{h}_v^{(-)}i\overrightarrow{D{\slash}}_{\bot}
(iv{\cdot}\overrightarrow{D}-2m)^{-1}
i\overrightarrow{D{\slash}}_{\bot}h_v^{(-)} 
\end{eqnarray}
as is well known in the literature (we omit hereafter
the small imaginary parts in the denominators). Besides we get for the 
particle-antiparticle sectors
\begin{eqnarray}
{\cal L}^{(-+)}\ =\ e^{-i2mv{\cdot}x} & \biggl[ & \overline{h}_v^{(-)}
i\overleftarrow{D}{\slash}_{\bot}
(-iv{\cdot}\overleftarrow{D}-2m)^{-1}(iv{\cdot}\overrightarrow{D})h_v^{(+)} \\
& + & \overline{h}_v^{(-)}i\overleftarrow{D}{\slash}_{\bot}
(-iv{\cdot}\overleftarrow{D}-2m)^{-1}
i\overrightarrow{D{\slash}}_{\bot}(iv{\cdot}\overrightarrow{D}+2m)^{-1}
i\overrightarrow{D{\slash}}_{\bot}h_v^{(+)}\ \biggr] \nonumber 
\end{eqnarray}
\begin{eqnarray}
{\cal L}^{(+-)}\ =\ e^{i2mv{\cdot}x} & \biggl[ & \overline{h}_v^{(+)}
i\overleftarrow{D}{\slash}_{\bot}
(-iv{\cdot}\overleftarrow{D}+2m)^{-1}(iv{\cdot}\overrightarrow{D})h_v^{(-)}  \\
& + & \overline{h}_v^{(+)}i\overleftarrow{D}{\slash}_{\bot}
(-iv{\cdot}\overleftarrow{D}+2m)^{-1}
i\overrightarrow{D}{\slash}_{\bot}(iv{\cdot}\overrightarrow{D}-2m)^{-1}
i\overrightarrow{D{\slash}}_{\bot}h_v^{(-)}\ \biggr] \nonumber
\end{eqnarray}
where the first piece (Eq. (24)) stands for heavy-quark pair annihilation
and the last piece (Eq. (25)) for heavy-quark pair creation 
\footnote{Superscript ${\lq\lq}(+)/(-)$" on the effective fields labels
the particle/antiparticle sector of the theory. Actually 
$\overline{h}_v^{(+)}\ (\overline{h}_v^{(-)})$ corresponds to
negative (positive) frequencies associated with 
creation (annihilation) operators of quarks (antiquarks).
This somewhat misleading notation is a consequence of the
former notation employed for the effective fields \cite{georgi}. 
In fact some extra ${\lq\lq}+/-$" signs should be added on the 
conjugate fields, 
i.e. $\overline{h}_v^{(+)-}$ and $\overline{h}_v^{(-)+}$, which
however will be omitted to shorten the notation.}.\par
Alternatively we can proceed directly from the Lagrangians (13-15) by
requiring the constraint on the effective fields given by
Eq. (16), obtaining in terms of $\lq\lq$large'' and
 $\lq\lq$small'' effective fields 
\par

\begin{equation}
{\cal L}^{(++)}\ =\ \overline{h}_v^{(+)}(iv{\cdot}\overrightarrow{D})h_v^{(+)}+
\overline{h}_v^{(+)}i\overrightarrow{D{\slash}}_{\bot}H_v^{(+)}
\vspace{0.1in}
\end{equation}

\begin{equation}
{\cal L}^{(--)}\ =\ 
-\overline{h}_v^{(-)}(iv{\cdot}\overrightarrow{D})h_v^{(-)}+
\overline{h}_v^{(-)}i\overrightarrow{D{\slash}}_{\bot}H_v^{(-)}
\end{equation}
and

\begin{equation}
{\cal L}^{(-+)}=e^{-i2mv{\cdot}x}\ 
[\ \overline{H}_v^{(-)}(iv{\cdot}\overrightarrow{D})h_v^{(+)}+
\overline{H}_v^{(-)}i\overrightarrow{D{\slash}}_{\bot}H_v^{(+)}\ ]  
\vspace{0.1in}
\end{equation}

\begin{equation}
{\cal L}^{(+-)}=e^{i2mv{\cdot}x}\ 
[\ -\overline{H}_v^{(+)}(iv{\cdot}\overrightarrow{D})h_v^{(-)}
+\overline{H}_v^{(+)}i\overrightarrow{D{\slash}}_{\bot}H_v^{(-)}\ ] 
\vspace{0.1in}
\end{equation}
recovering previous expressions (22-25) by making subsequent use of
Eqs. (18-21).

Let us note that, at first sight, one might think that the rapidly oscillating
exponential in Eqs. (28-29) would make both  ${\cal L}^{(-+)}$ and 
${\cal L}^{(+-)}$ pieces vanish once integrated over
all velocities $v$ according to the most general Lagrangian \cite{georgi}.
However, notice that actually this is not the case for momenta of order 
$2mv$ of the
gluonic field present in the covariant derivative. In fact, only
such high-energy modes would survive corresponding to the physical
situation on which we are focusing, i.e. heavy quark-antiquark pair
annihilation and \vspace{0.2in} creation.

\subsection{Symmetrized Lagrangian}
Had we started from the very beginning with the symmetrized QCD Lagrangian 
in Eq. (8)

\begin{equation}
{\cal L}\ =\ \overline{Q}\ 
\biggl(i\frac{\overleftrightarrow{D{\slash}}}{2}-m\biggr)\ Q
\end{equation}
where
\[ \overleftrightarrow{D}\ =\ \overrightarrow{D}\ -\ 
\overleftarrow{D} \]
then Eqs. (14) and  (15) should be replaced by 

\begin{eqnarray}
{\cal L}^{(-+)}\ =\ e^{-i2mv{\cdot}x} & \biggl[ & \overline{H}_v^{(-)}
\biggl(\ iv{\cdot}\frac{\overleftrightarrow{D}}{2}-m\biggr)h_v^{(+)}\ 
-\ \overline{h}_v^{(-)}\biggl(iv{\cdot}\frac{\overleftrightarrow{D}}{2}+
m\biggr)H_v^{(+)} \\
& + & 
\overline{h}_v^{(-)}i\frac{\overleftrightarrow{D{\slash}}_{\bot}}{2}h_v^{(+)}
\ +\ 
\overline{H}_v^{(-)}i\frac{\overleftrightarrow{D{\slash}}_{\bot}}{2}
H_v^{(+)}\ \biggr] \nonumber 
\vspace{0.1in}
\end{eqnarray}
and

\begin{eqnarray}
{\cal L}^{(+-)}\ =\  e^{i2mv{\cdot}x} & \biggl[\ & -\overline{H}_v^{(+)}
\biggl(iv{\cdot}\frac{\overleftrightarrow{D}}{2}+m\biggr)h_v^{(-)}\ 
+\ \overline{h}_v^{(+)}\biggl(iv{\cdot}
\frac{\overleftrightarrow{D}}{2}-m\biggr)H_v^{(-)} \\
& + &
\overline{h}_v^{(+)}i\frac{\overleftrightarrow{D{\slash}}_{\bot}}{2}h_v^{(-)}+
\overline{H}_v^{(+)}i\frac{\overleftrightarrow{D{\slash}}_{\bot}}{2}
H_v^{(-)}\ \biggr] \nonumber 
\end{eqnarray}
yielding in terms of the $h_v^{(\pm)}$ fields

\begin{eqnarray}
{\cal L}^{(-+)}\ =\ e^{-i2mv{\cdot}x} & \biggl[ & \overline{h}_v^{(-)}
i\overleftarrow{D}{\slash}_{\bot}
(-iv{\cdot}\overleftarrow{D}-2m)^{-1}(iv{\cdot}\overrightarrow{D})
h_v^{(+)} \nonumber \\
& + & \overline{h}_v^{(-)}
(iv{\cdot}\overleftarrow{D})(iv{\cdot}\overrightarrow{D}+2m)^{-1}
i\overrightarrow{D{\slash}}_{\bot}h_v^{(+)} \\
& + & \overline{h}_v^{(-)}i\overleftarrow{D{\slash}}_{\bot}
(-iv{\cdot}\overleftarrow{D}-2m)^{-1}
i\overleftrightarrow{D{\slash}}_{\bot}
(iv{\cdot}\overrightarrow{D}+2m)^{-1}
i\overrightarrow{D{\slash}}_{\bot}h_v^{(+)}\ \biggr] \nonumber 
\end{eqnarray}

\begin{eqnarray}
{\cal L}^{(+-)}\ =\ e^{i2mv{\cdot}x} & \biggl[ & \overline{h}_v^{(+)}
i\overleftarrow{D}{\slash}_{\bot}
(-iv{\cdot}\overleftarrow{D}+2m)^{-1}(iv{\cdot}
\overrightarrow{D})h_v^{(-)}  \nonumber \\
& + & \overline{h}_v^{(+)}
(iv{\cdot}\overleftarrow{D})(iv{\cdot}\overrightarrow{D}-2m)^{-1}
i\overrightarrow{D{\slash}}_{\bot}h_v^{(-)} \\
& + & \overline{h}_v^{(+)}i\overleftarrow{D{\slash}}_{\bot}
(-iv{\cdot}\overleftarrow{D}+2m)^{-1}
i\overleftrightarrow{D{\slash}}_{\bot}
(iv{\cdot}\overrightarrow{D}-2m)^{-1}
i\overrightarrow{D{\slash}}_{\bot}h_v^{(-)}\ \biggr] \nonumber
\end{eqnarray}
having dropped an overall $1/2$ factor in the two last 
expressions constituting the generalization of Eqs. (24-25).
\par
\subsection{Troubles with the $1/m$ expansion in the mixed sectors}

Notice that either piece of the effective 
Lagrangian mixing heavy quarks and antiquarks cannot be 
straightforwardly expanded in powers of the reciprocal of the 
heavy-quark mass. This can be seen assuming that
the gluon field has an $x$-dependence of the type \par

\[ A_{\mu}^{\mp}(x)\ =\ \epsilon_{\mu}^{(\ast})\ e^{{\pm}i(2mv+k){\cdot}x} \]
for annihilation/creation processes of heavy quarks almost on-shell.
Thus, from Eq. (16) we may write 
\begin{eqnarray}
H_v^{(\pm)} & = & \frac{1}{{\pm}iv{\cdot}\overrightarrow{D}+2m}
i\overrightarrow{D{\slash}}_{\bot}h_v^{(\pm)}
\nonumber \\
& = & \frac{1}{2m}\biggl(1{\mp}\frac{(iv{\cdot}\overrightarrow{D})}{2m}+
\frac{(iv{\cdot}\overrightarrow{D})^2}{4m^2}{\mp}
...\biggr)\ (i\overrightarrow{\partial}{\slash}_{\bot}
+gA{\slash}_{\bot}^{\mp})h_v^{(\pm)}  
\end{eqnarray}

\def\slash{\!\!\!/}

One may easily see that the above Taylor expansions 
are not convergent because of the strong $x$ dependence
of the $A{\slash}_{\bot}^{\mp}h_v^{(\pm)}$ term. In particular the
derivatives acting on $A{\slash}_{\bot}^{\mp}h_v^{(\pm)}$ bring
mass powers exactly cancelling the mass dependence of the denominators
in the successive terms of the expansion. In other words, the
rapidly oscillating gluonic field due to its large momentum spoils the 
power expansion in (35) apparently  making useless the effective theory 
framework in this case. This is in contradistinction to the scattering
of a heavy quark or a heavy antiquark by a soft gluon where the above 
expansion does make sense (no mass powers then appearing when acting the
derivatives on the exponentials leading to well-poised $1/m$ series) 
as usually employed in HQET. \par
In sum, Lagrangians ${\cal L}^{(++)}$ and ${\cal L}^{(--)}$
can be expanded in terms of local operators whereas
the  ${\cal L}^{(-+)}$ and ${\cal L}^{(+-)}$ pieces, as shown in
Eqs. (24-25) or (33-34), are
unexpansible non-local \vspace{0.2in} Lagrangians.\par

\section{A $1/m$ expansion appropriate for   
 ${\cal L}^{(-+)}$ and ${\cal L}^{(+-)}$}
Although  expressions (24-25) can be considered as formally correct, we have
remarked in the last section that, in fact,  they are not appropriate to deal
with heavy quark pair production or annihilation. Essentially
this is because in using the equations of motion we
assumed heavy quarks/antiquarks to be in the background of a
strong color field, leading to non-convergent expansion
series for the effective fields.\par
Therefore in this section we shall require
that the effective fields satisfy, instead, the corresponding equations of
motion for (almost) free quarks and antiquarks. In other words, effective
fields should obey the classical equations of motion derived
from the ${\cal L}^{(++)}$ and ${\cal L}^{(--)}$ pieces
though used for the ${\cal L}^{(-+)}$ or ${\cal L}^{(+-)}$
sectors of the theory. In the following we shall focus
on the heavy quark-antiquark pair annihilation into a gluon as shown
in figure 1. For simplicity we assume both heavy quark and antiquark
on-shell but only the former one with a residual momentum $k$ satisfying
the condition (1). Indeed if one is interested in doubly heavy hadrons
such as heavy quarkonium, initial-state heavy quarks are almost
on shell with small residual momentum, of order 
\vspace{0.2in} $m{\alpha}_s<<m$.

\begin{figure}[htb]
\centerline{
\epsfig{figure=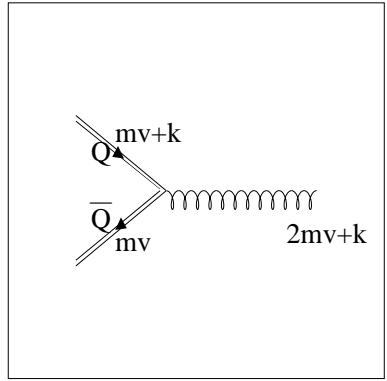,height=5.cm,width=5.cm}}
\caption{\small{Heavy quark-antiquark pair annihilation into
a gluon of momentum $2mv+k$. For simplicity  the heavy quark 
and antiquark are assumed to carry momenta $mv+k$ and $mv$ respectively. Such
process could be described meaningfully by HQET if $k$ is small w.r.t
 the heavy quark mass $m$, i.e. the relative velocity of
both heavy quarks is small. The same statement is valid for heavy quark
pair production near theshold.}}
\end{figure}

\par

Assuming the same conditions as in figure 1, we can write
for almost free heavy quarks in the initial- or final-state

\def\slash{\!\!\!/}

\begin{eqnarray}
H_v^{(+)} & = & (iv{\cdot}\overrightarrow{\partial}+2m-i\epsilon)^{-1}
i\overrightarrow{\partial{\slash}}_{\bot}h_v^{(+)} \\
H_v^{(-)} & = & (-iv{\cdot}\overrightarrow{\partial}+2m-i\epsilon)^{-1}
i\overrightarrow{\partial{\slash}}_{\bot}h_v^{(-)}
\end{eqnarray}
and for the conjugate fields
\begin{eqnarray}
\overline{H}_v^{(+)} & = & 
\overline{h}_v^{(+)}i\overleftarrow{\partial{\slash}}_{\bot}
(iv{\cdot}\overleftarrow{\partial}-2m+i\epsilon)^{-1} \\
\overline{H}_v^{(-)} & = & 
\overline{h}_v^{(-)}i\overleftarrow{\partial{\slash}}_{\bot}
(-iv{\cdot}\overleftarrow{\partial}-2m+i\epsilon)^{-1}
\end{eqnarray}

Neglecting the soft gluon interaction among heavy quarks amounts to 
describe them as plane waves, i.e. actually no bound states 
as a first approximation \cite{hussain}.

Expanding and retaining the first non-null terms in Eqs. (36) and (39)
one easily gets

\begin{equation}
H_v^{(+)}\ =\ \frac{i\overrightarrow{\partial{\slash}}_{\bot}}{2m}\ h_v^{(+)}
\end{equation}
\begin{equation}
\overline{H}_v^{(-)}\ =\ -\ \overline{h}_v^{(-)}\ 
\frac{i\overleftarrow{\partial{\slash}}_{\bot}}{2m}
\end{equation}

In momentum space ($k$ representing the residual momentum of
the heavy quark w.r.t. the heavy antiquark) one may write to 
the same order in the $1/m$ expansion

\begin{equation}
u\ =\ \biggl(1\ +\ \frac{k{\slash}_{\bot}}{2m}\biggr)\ u_h\ e^{-ik{\cdot}x}
\end{equation}
where $u_h$ stands for the HQET spinor and $u$ for the full QCD
spinor. On the other hand
\begin{equation}
\overline{v}\ =\ \overline{v}_h
\end{equation}
where $v_h$ stands for the HQET anti-spinor for a heavy antiquark
without residual momentum and $v$ denotes the full QCD anti-spinor.

Since $\overline{H}^{(-)}{\equiv}0$ in this case, Eq. (14) simplifies in 
this particular case to

\def\slash{\!\!\!\!/}

\begin{equation}
{\cal L}^{(-+)}=e^{-i2mv{\cdot}x}\ 
[\ \overline{h}_v^{(-)}i\overrightarrow{D{\slash}}_{\bot}h_v^{(+)}
-\overline{h}_v^{(-)}(iv{\cdot}\overrightarrow{D}+2m)H_v^{(+)}\ ]
\end{equation}

It is convenient to emphasize that the covariant derivative in the
above expression contains the strong gluonic field
while the $H_v^{(+)}$ field satisfies Eq. (36). Thus
the  effective annihilation vertex into
a final gluon can be read off as
\begin{equation}
i{\Gamma}_{eff}^{\mu}\ =\ igT_a\ \biggl[\ 
{\gamma}^{\mu}\ -\ i\frac{{\sigma}^{\mu\nu}k_{\nu}}{2m}\ +\  
\frac{k^2}{4m^2}{\gamma}^{\mu}\ +\ ...\ 
\biggr]
\end{equation}
\par

\def\slash{\!\!\!/}

Indeed, from the terms of (14) coupling to one gluon we may
write the effective vertex as
\begin{equation}
i{\Gamma}_{eff}^{\mu}\ =\ igT_a\ \biggl[\ 
{\gamma}^{\mu}\ -\ \frac{k{\slash}}{2m}v^{\mu}\ +\  ...\ 
\biggr]
\end{equation}
and then using the relation (A.2) written as
\begin{equation}
P_-\ (v^{\mu}{\gamma}^{\nu})\ P_+\ =\ P_-\ (i{\sigma}^{\mu\nu}\ +\ 
{\gamma}^{\mu}v^{\nu})\ P_+
\end{equation}
we obtain (45) if the on-shell condition for the heavy quark is fulfilled, 
i.e.
\[ v{\cdot}k\ =\ -\frac{k^2}{2m} \]

\par
In fact  the same result could be directly derived from full 
(tree-level) QCD by
making the appropriate transformations in the Feynman amplitude
of the vertex itself, namely

\begin{eqnarray}
igT_a\ \overline{v}\ {\gamma}^{\mu}\ u & {\rightarrow} & 
igT_a\ \overline{v}_h\ {\gamma}^{\mu} 
\biggl[\ 1\ +\ \frac{k{\slash}_{\bot}}{2m}\ + ... \biggr]\ u_h
\nonumber \\
& = & igT_a\ 
\overline{v}_h\ {\gamma}^{\mu} \biggl[\ 1\ +\ \frac{k{\slash}}{2m}\ 
-\ \frac{(v{\cdot}k)}{2m}\ +\ ...\ \biggr]\ u_h 
\end{eqnarray}
recovering the result expressed in Eq. (45) since                                     
\[ P_-\ \biggl(\frac{{\gamma}^{\mu}k{\slash}}{2m}\biggr)\ P_+\ =\ 
P_-\ \biggl(-i\frac{{\sigma}^{\mu\nu}k_{\nu}}{2m}\biggr)\ P_+ \]

The above exercise shows that indeed the effective theory and
full QCD lead to the same annihilation vertex at the same
order in the $1/m$ \vspace{0.2in} expansion.

\section{Summary and last remarks}

We have derived a complete tree-level heavy-quark effective Lagrangian in 
terms of the effective fields $h_v^{(\pm)}$ keeping the 
particle-antiparticle mixed pieces allowing for heavy quark-antiquark 
pair annihilation/creation processes. Let us note that such pieces are not
generally shown in similar developments in the literature. 
\par
Indeed, it may seem quite striking that a low-energy effective theory 
could be appropriate to deal
with hard processes such as $Q\overline{Q}$ annihilation or creation.
The keypoint is that assuming a kinematic regime where heavy quarks/antiquarks
are almost on-shell and moving with small relative velocity, one can
factor out the strong momentum dependence associated with the heavy
quark mass both at the initial and final state (i.e. including also
the strong gluonic field) so that a description based on the low frequency
modes of the fields still makes sense. Such a kinematic regime is well 
matched by heavy quarkonium states and by colored
intermediate bound states in the so-called color-octet model recently
introduced \cite{fleming}  to account for high rates of 
heavy resonance hadroproduction
at the Fermilab Tevatron \cite{fermi}. \par 
However, a systematic inverse heavy-quark mass expansion of
the mixed pieces $L^{(-+)}$ and $L^{(+-)}$ is not permitted
if the classical equations of motion for heavy quarks in the background of a 
strong gluonic field are employed. This is because 
Eq. (35), where effective fields $H_v^{(\pm)}$ are written
in terms of $h_v^{(\pm)}$, is not a convergent expression due to the 
momentum dependence of the strong gluonic field, of order of twice 
the heavy quark mass. (We might say in a loose way that the $H_v^{(\pm)}$ 
fields are not $\lq\lq$small'' anymore.)
As a consequence the Lagrangian pieces $L^{(-+)}$ and $L^{(+-)}$ as
expressed in Eqs. (28-29) or (33-34) are not
expansible in powers of the reciprocal of the heavy quark
mass. Moreover, let us note that in using such equations of motion
heavy quarks would be assumed in the background of a strong color
field. However, we are rather interested in a kinematic regime
where initial- or final-state heavy quarks interact softly among themselves 
or with light constituents.\par
Therefore, if the classical equations of motion deriving from the
$L^{(++)}$ and $L^{(--)}$ sectors are used to remove the
unwanted degrees of freedom in the $L^{(-+)}$ and $L^{(+-)}$
sectors, then convergent and meaningful powers series are obtained.
In particular, assuming an annihilation process with initial-state 
quarks (or final-state quarks for a creation process) satisfying 
the Dirac equation of motion for free fermions, the vertex
Feynman rule is easily obtained, in accordance with full
QCD vertex expanded to the same $1/m$ order.

In summary, we conclude that heavy quark-antiquark pair annihilation
and creation can be meaningfully described starting from a tree-level
HQET Lagrangian for special kinematic regimes of current interest in
heavy flavor \vspace{0.1in} physics.

\subsection*{\bf Acknowledgments}
We thank V. Gim\'enez, A. Pineda and J. Soto
for reading the manuscript and critical \vspace{0.1in} comments.

\newpage

\appendix\renewcommand{\theequation}{\thesection.\arabic{equation}}
\section*{Appendices} 
\section{Useful spinorology}
\setcounter{equation}{0}

\def\slash{\!\!\!/}

In this appendix we gather some well-known simple formulas together
with some others not so commonly found in the literature, exhaustively
employed throughout this work
\par
\[
v{\slash}h_v^{(+)}=h_v^{(+)}\ ;\ v{\slash}H_v^{(+)}= -H_v^{(+)}\ ;\ 
v{\slash}h_v^{(-)}=-h_v^{(-)}\ ;\ v{\slash}H_v^{(-)}=H_v^{(-)} 
\]
Then, the projectors $P_{\pm}=(1{\pm}v{\slash})/2$ acting on the fields yield
\begin{eqnarray}
& & P_+\ h_v^{(+)}=h_v^{(+)}\ \ ;\ \ 
P_-\ h_v^{(+)}=0\ \ ;\ \ P_-\ H_v^{(+)}=H_v^{(+)}\ \ ;\ \
P_+\ H_v^{(+)}=0 \nonumber \\ 
& & P_-\ h_v^{(-)}=h_v^{(-)}\ \ ;\ \  
P_+\ h_v^{(-)}=0\ \ ;\ \ P_+\ H_v^{(-)}=H_v^{(-)}\ \ ;\ \ 
P_-\ H_v^{(-)}=0 \nonumber 
\end{eqnarray}
and analogously for the conjugate fields $\overline{h}_v^{(+)}$, 
$\overline{h}_v^{(-)}$, $\overline{H}_v^{(+)}$ and $\overline{H}_v^{(-)}$.
Therefore orthogonality implies
\[
\overline{h}_v^{(-)}h_v^{(+)}\ =\ \overline{h}_v^{(-)}H_v^{(-)}\ =\ 
\overline{H}_v^{(-)}h_v^{(-)}\ =\
\overline{H}_v^{(-)}H_v^{(+)}\ =\ 0
\]
but
\[
\overline{h}_v^{(-)}H_v^{(+)}\ {\neq}\ 0\ \ ;\ 
\overline{h}_v^{(-)}{\gamma}^{\mu}\ h_v^{(+)}\ {\neq}\ 0 
\]
and so on.\par

In the particle-particle sector of the effective theory, it is also
well-known that the vector current can be expressed as
\[ \overline{h}_v^{(+)}\ {\gamma}^{\mu}\ h_v^{(+)}\ =\  
\overline{h}_v^{(+)}\ v^{\mu}\ h_v^{(+)} \]
\par
On the other hand, since
\begin{equation}
P_-\ {\gamma}^{\mu}{\gamma}^{\nu}\ P_+\ =\  
P_-\ ({\gamma}^{\mu}v^{\nu}\ -\
v^{\mu}{\gamma}^{\nu})\ P_+
\end{equation}

an equivalent relation in the particle-antiparticle sector reads
\[ \overline{h}_v^{(-)}\ {\gamma}^{\mu}{\gamma}^{\nu}\ h_v^{(+)}\ =\  
\overline{h}_v^{(-)}\ {\gamma}^{\mu}v^{\nu}\ h_v^{(+)}\ -\
\overline{h}_v^{(-)}\ v^{\mu}{\gamma}^{\nu}\ h_v^{(+)} \]
This implies that
\begin{equation}
P_-\ (-i\sigma^{\mu\nu})\ P_+\ =\ P_-\ 
({\gamma}^{\mu}v^{\nu}-v^{\mu}{\gamma}^{\nu})\ P_+
\end{equation}
and
\begin{equation}
P_-\ (-i\sigma^{\mu\nu})\ P_+\ =\ P_-\ 
({\gamma}^{\mu}{\gamma}^{\nu})\ P_+
\end{equation}
and thus
\begin{equation}
\overline{h}_v^{(-)}\ (-i\sigma^{\mu\nu})\ h_v^{(+)}\ =\ 
\overline{h}_v^{(-)}\ {\gamma}^{\mu}v^{\nu}\ h_v^{(+)}\ -\                     
\overline{h}_v^{(-)}\ v^{\mu}{\gamma}^{\nu}\ h_v^{(+)}
\end{equation}
Hence
\begin{equation}
\overline{h}_v^{(-)}\ (-i\sigma^{\mu\nu}v_{\nu})\ h_v^{(+)}\ =\ 
\overline{h}_v^{(-)}\gamma^{\mu}\ h_v^{(+)}
\end{equation} 

\par
In the particle-particle sector odd combinations of gamma matrices can be 
simplified by replacing some ${\gamma}$ matrices by $v$'s, whereas in
the particle-antiparticle sector
even combinations lead to the same kind of simplification. For example,
\begin{equation} 
P_+\ {\gamma}^{\mu}{\gamma}^{\nu}{\gamma}^{\alpha}\ P_+\ =\  
P_+\ (v^{\mu}{\gamma}^{\nu}{\gamma}^{\alpha}\ -\
{\gamma}^{\mu}v^{\nu}{\gamma}^{\alpha}\ +\ 
{\gamma}^{\mu}{\gamma}^{\nu}v^{\alpha})\ P_+ 
\end{equation}
Permutation of the projectors $P_{\pm}$ in all above
expressions implies the overall substitution $v{\rightarrow}-v$. 
Moreover, similar relations are satisfied if the $\gamma$ matrices are
sandwiched between heavy quark spinors and/or anti-spinors
obeying $v{\slash}u_h=u_h$ and $v{\slash}v_h=-v_h$.

\end{document}